\documentclass[runningheads]{llncs}
\usepackage{graphicx}
\usepackage{booktabs} 
\usepackage{amssymb}
\usepackage{xcolor}
\usepackage{amsmath}
\usepackage{colortbl}
\usepackage{amssymb}
\usepackage{marvosym}
\usepackage{bm}
\definecolor{right}{rgb}{.3,.7,.1}
\definecolor{frenchblue}{rgb}{0.0, 0.45, 0.73}

\begin{document}

%
\title{Prompting Whole Slide Image Based Genetic Biomarker Prediction}
%
\author{Ling Zhang\inst{1} \and
Boxiang Yun\inst{1} \and Xingran Xie\inst{1} \and Qingli Li\inst{1} \and Xinxing Li\inst{2} \and \\Yan Wang\inst{1}\textsuperscript{(\Letter)}}

\institute{{Shanghai Key Laboratory of Multidimensional Information Processing, \\East China Normal University, Shanghai, China\and
Department of Gastrointestinal Surgery, \\ Tongji Hospital Medical College of Tongji University}\\
\email{51265904017@stu.ecnu.edu.cn, 52265904012@stu.ecnu.edu.cn, 51215904112@stu.ecnu.edu.cn, qlli@cs.ecnu.edu.cn, ahtxxxx2015@163.com, ywang@cee.ecnu.edu.cn}
}
%
\maketitle              
\begin{abstract}
Prediction of genetic biomarkers, \emph{e.g.}, microsatellite instability and BRAF in colorectal cancer is crucial for clinical decision making. 
In this paper, we propose a whole slide image (WSI) based genetic biomarker prediction method via prompting techniques. 
Our work aims at addressing the following challenges: 
(1) extracting foreground instances related to genetic biomarkers from gigapixel WSIs, and (2) the interaction among the fine-grained pathological components in WSIs. 
Specifically, we leverage large language models to generate medical prompts that serve as prior knowledge in extracting instances associated with genetic biomarkers. We adopt a coarse-to-fine approach to mine biomarker information within the tumor microenvironment. This involves extracting instances related to genetic biomarkers using coarse medical prior knowledge, grouping pathology instances into fine-grained pathological components and mining their interactions. 
Experimental results on two colorectal cancer datasets show the superiority of our method, achieving 91.49\% in AUC for MSI classification. The analysis further shows the clinical interpretability of our method. Code is publicly available at https://github.com/DeepMed-Lab-ECNU/PromptBio.

\keywords{Genetic biomarker prediction  \and Whole slide image  \and Text prompt.}
\end{abstract}

\section{Introduction}
Colorectal cancer (CRC) is the second leading cause of cancer-related deaths worldwide \cite{sung2021global}. The evaluation of genetic biomarkers is crucial for CRC diagnosis and prognosis, such as microsatellite instability (MSI) and mutations in the BRAF, since these biomarkers can identify CRC patients with different treatment response and prognosis \cite{koncina2020prognostic,caputo2019braf,chang2018microsatellite}. MSI has been officially adopted as global classification standards for the diagnosis and treatment of CRC in the National Comprehensive Cancer Network guidelines \cite{ajani2023esophageal}. However, the method of testing genetic biomarkers is time-consuming and expensive, such as immunohistochemistry, polymerase chain reaction and next-generation-sequencing \cite{dedeurwaerdere2021comparison}. Due to diagnostic needs, the pathology whole slide images (WSIs) stained with hematoxylin and eosin (H\&E)  are routinely available for all CRC patients and previous works \cite{kather2019deep,shimada2021histopathological} suggested that genetic alterations are expressed in digital pathology WSIs. Therefore, automatically predicting genetic biomakers from WSIs is feasible and highly demanded in clinical practice.

Genetic biomarker prediction based on gigapixel WSIs can be cast as a multiple instance learning (MIL) problem, where WSIs are bags and patches cropped from WSIs are instances, aiming to predict a bag label. However, learning genetic biomarker representations based on bags is challenging, because it's hard to extract foreground instances related to genetic biomarkers from a WSI. Moreover, the tumor microenvironment is a complex network. Various pathological component interactions exist between immune cells and non-cellular components such as the extracellular matrix, exosomes and interleukins \cite{bozyk2022tumor}, which 
may be relevant to genetic biomarkers. Therefore, how to  extract foreground instances related to genetic biomarkers and mine the interaction among pathological components in the the tumor microenvironment is crucial for genetic biomarker prediction.

Specifically, to our knowledge, most WSI based genetic biomarker prediction methods, such as MOMA \cite{tsai2023histopathology} and MSIntuit \cite{saillard2023validation}, indiscriminately input all instances into the network, which bring a large number of irrelevant instances.  Many other WSI classification methods can be extended to predict genetic biomarkers. Among these, some methods have implemented learnable or non-learnable instance selection. For example, VIBFT \cite{li2023task} introduces a variational information bottleneck  to find the minimal sufficient statistics of WSIs, and MILBooster \cite{qu2023boosting} proposes a bag filter that  increases the positive instance ratio of positive bags using K-means algorithm. But these selection methods lack interpretable medical justification. Even with a reduced number of instances, they may still lose key tissue regions relevant to genetic biomarker. Moreover, most methods \cite{Ilse_Tomczak_Welling_2018,li2021dual}  treat pathology instances as independent entities and overlook their interaction. Recent studies \cite{shao2021transmil,li2021dt} have used Vision Transformer frameworks to model the relationships between instances, but they neglect the interaction among different pathological components. HGT \cite{hou2023multi} attempts to explore the interaction among different pathological components in WSIs, but their construction of pathological components relies on  superpixel rather than actual pathological component distribution.  TOP \cite{qu2024rise} models different prototypes guided by pathology language prior knowledge and the pathology language prior knowledge is flexible and adaptable, but it aggregates instance features into a bag feature through a simple weighting method without exploring the interaction among different prototypes.

To this end, we propose a task-specific framework dubbed Prompting whole slide image-based genetic Biomarker prediction (PromptBio), consisting of three modules, which leverages the capabilities of the large language model (LLM) to generate pathology text prompts, and uses biomarker-associated medical prior knowledge to guide the pathological component extraction and feature fusion. 
Pathological components of diverse microenvironments range from fine-grained (\emph{e.g.}, lymphocyte infiltration, inflammatory reaction, \emph{etc.}) to coarse-grained (\emph{e.g.}, cancer-associated stroma, epithelium tissue, \emph{etc}.).
Among various coarse-grained pathological components, cancer-associated stroma is proven as the \textbf{key contributor} to the tumor microenvironment \cite{bussard2016tumor,brown2022cancer} and is highly associated with consensus molecular subtypes \cite{becht2016immune}. We regard cancer-associated stroma as the tissue of our interest and use it to discard irrelevant information in a WSI. Thus, we first propose a coarse-grained pathological instance selection module to extract foreground instances belonging to cancer-associated stroma from a gigapixel WSI. This process is guided by coarse-grained medical prompts. Next, we focus on learning the distribution and interaction among fine-grained pathological components in cancer-associated stroma, to extract comprehensive and in-depth biomarker-associated features. Two modules, \emph{i.e.}, fine-grained pathological component grouping and interaction mining, are proposed. In the grouping module, we prompt LLM to generate biomarker-associated descriptions for fine-grained pathological components, which identify multiple semantic types from the WSIs. We design a grouping strategy to aggregate instances of each type and gather their distribution. Prompting techniques ensure the accuracy of the tumor microenvironment modeling. Furthermore, the hierarchical contextual interaction of intra- and inter-pathological components are extracted and fused via our interaction mining module, enhancing the interpretability of the tumor microenvironment modeling. Extensive experiments show our method outperforms all state-of-the-arts by a large margin, with over 5\% improvement in AUC for CRC MSI prediction on TCGA dataset. Additionally, our method shows clinical interpretability, providing potential for future clinical applications.

\begin{figure}[t]
\includegraphics[width=\textwidth]{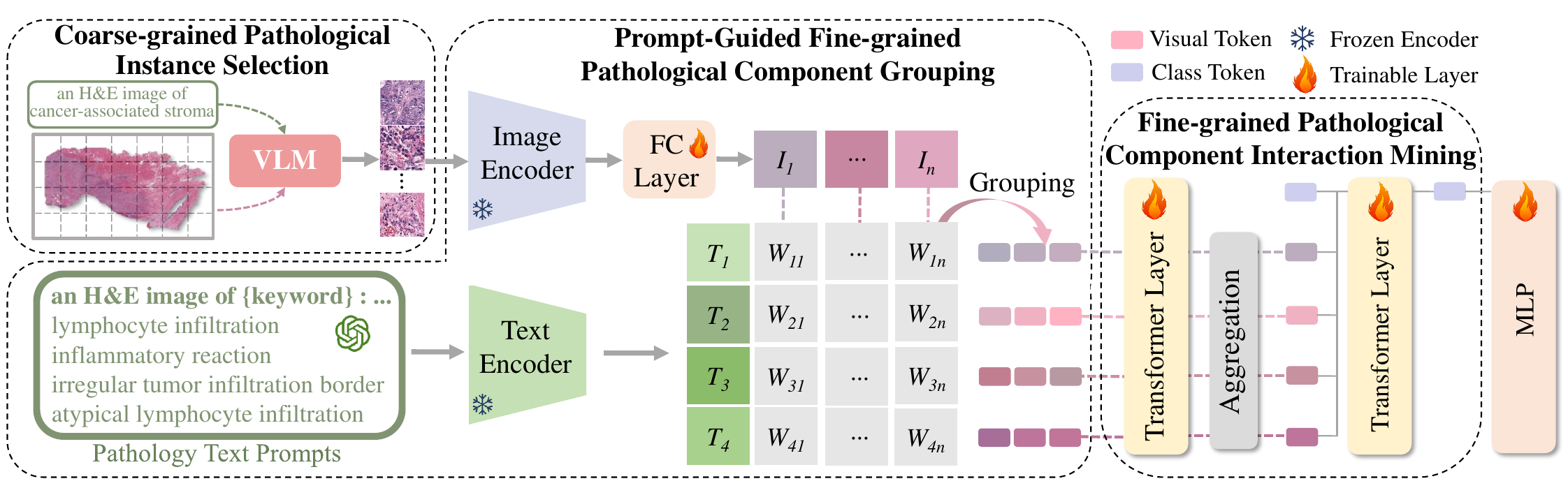}
\vspace{-1em}
\caption{Illustration of PromptBio model. The overall framework consists of three parts: 1) coarse-grained pathological instance selection module, 2) prompt-guided fine-grained pathological component grouping module, 3) fine-grained pathological component interaction mining module. Given a dataset  $\mathcal{D}$ consisting of $N$ pathology WSIs, our PromptBio first performs coarse-grained pathological instance selection to extract the instances belonging to cancer-associated stroma. Then our PromptBio performs fine-grained pathological component grouping on extracted instances. The grouping is guided by pathology text prompts of cancer-associated stroma with MSI. Finally, our PromptBio performs fine-grained pathological component interaction mining. } 
\vspace{-1em}
\label{fig1}
\end{figure}
\section{Method}
Given a dataset $\mathcal{D}$ consisting of $N$ pathology WSIs, pathological components of diverse microenvironments in WSIs range from fine-grained (\emph{e.g.}, lymphocyte infiltration, \emph{etc.}) to coarse-grained (cancer-associated stroma, \emph{etc}.). Prompting Whole Slice Image Based Genetic Biomarker Prediction (PromptBio) first performs coarse-grained pathological instance selection based on medical prior knowledge to extract the instances belonging to cancer-associated stroma. Considering that the tumor microenvironment is a complex network of various pathological components, PromptBio performs fine-grained pathological component grouping on extracted instances. The grouping is guided by pathology text prompts of cancer-associated stroma with MSI. Finally, PromptBio performs fine-grained pathological component interaction mining. As shown in Fig.\ref{fig1}, the overall framework consists of three modules: 1) coarse-grained pathological instance selection (Sec. 2.1), 2) prompt-guided fine-grained pathological component grouping (Sec. 2.2) and 3) fine-grained pathological component interaction mining (Sec. 2.3). Next, we will introduce each part in detail.
\subsection{Coarse-grained Pathological Instance Selection}
To  discard abundant irrelevant information in a WSI, we select the instances related to genetic biomarkers based on medical prior knowledge. Among various coarse-grained pathological components, cancer-associated stroma is proven as the key contributor to the tumor microenvironment \cite{bussard2016tumor,brown2022cancer,becht2016immune}. We regard cancer-associated stroma as the tissue of our interest and use it to discard abundant irrelevant information in a WSI. We utilize PLIP \cite{huang2023visual} for coarse-grained pathological instance classification of CRC WSIs. PLIP is a large vision-language model (VLM) for pathology image analysis that allows zero-shot classification of tissues on WSIs, where the text encoder yields text embeddings of medical descriptions. An H\&E-stained WSI contains a large number of irrelevant background regions, and we use the image preprocessing algorithm in CLAM \cite{lu2021data} to detect the tissue regions in the WSI and crop the tissue regions into non-overlapping 512 × 512 pixel-sized patches. Then we input the text ``an H\&E image of \{keywords\}'' into the frozen text encoder of PLIP and input the instances of the WSIs into the image encoder of PLIP. The keywords correspond to nine tissue classes \cite{huang2023visual} including cancer-associated stroma. Finally, we extract the instances belonging to cancer-associated stroma.
\subsection{Prompt-Guided Fine-grained Pathological Component Grouping} 
To learn the distribution of fine-grained pathological components in cancer-associated stroma and extract more comprehensive and in-depth biomarker-associated features, we group selected instances into fine-grained pathological components based on pathology text prompts. We use the prompt template
“If a colorectal cancer patient is in an MSI (Microsatellite Instability) status, what are the characteristics of the cancer-associated stroma areas in the WSIs may exhibit?” for guiding GPT-4 to generate the pathology descriptions. The generated descriptions include four types: lymphocyte infiltration, inflammatory reaction, irregular tumor infiltration border, and atypical lymphocyte infiltration. Each type has a specific description. We input the four pathology descriptions into the text encoder of PLIP to obtain the text embeddings $\bm{T} \in \mathbb{R}^{4 \times d}$. After patch selection, we follow the method in  IBMIL \cite{lin2023interventional} to extract the instance features and then use a fully connected layer to map the instance features to $d$ dimensions $\bm{I} \in \mathbb{R}^{n \times d}$, where $n$ is the number of instances selected in each bag. We compute the cosine similarity between the instance features and each text embedding:
\begin{equation}
\bm{W}_i(\bm{I})=\text{sim}(\bm{I},{\bm{T}_i}^\top) , i=1,2,3,4,
\end{equation}
where sim(\textit{·}\textit{ , }\textit{·}) refers to cosine similarity, $\bm{T}_i$ means the $i$th text embedding of the four  components and $\bm{W}_i(\bm{I}) \in \mathbb{R}^{n \times 1}$ . 

Then we select the top-$k$ image features for each pathological component based on similarity ranking, where $k=n\cdot \beta$ and $\beta$ is the selection ratio. In this way, we obtain four fine-grained pathological components $\bm{\mathcal{C}}_i \in \mathbb{R}^{k \times d}$ and corresponding instance features $\bm{\mathcal{C}}_{i j} \in \mathbb{R}^{1 \times d}$, where $\bm{\mathcal{C}}_i$ means the $i$th component in the four components with pathology semantics. After grouping, many irrelevant instances are reduced. $\bm{\mathcal{C}}_i$ is computed by:
\begin{equation}
\bm{\mathcal{C}}_i=\Phi\left(\bm{W}_i\left(\bm{I}\right),k\right), i=1,2,3,4,
\end{equation}
where $\Phi(\bm{W}_i(\bm{I}),k)$ refers to the $k$ out of $n$ instances in $\bm{I}$ that maximize $\bm{W}_i\left(\bm{I}\right)$.
\subsection{Fine-grained Pathological Component Interaction Mining} 
To mine the interaction among fine-grained pathological components, we separately input the instance features of each pathological component into the transformer layer to model the interaction of each instance within a pathological component, and separately aggregate the instance features of each pathological component. The feature $\bm{\mathcal{C}}^{mean}_i\in \mathbb{R}^{1 \times d}$ of the $i$th pathological component is: 
\begin{equation}
\bm{\mathcal{C}}^{mean}_i=\frac{1}{k}\left(\sum_j^k \bm{\mathcal{C}}_{i j}\right), i=1,2,3,4.
\end{equation}
Then, we concatenate component features $\bm{\mathcal{C}}^{mean}=\text{Concat}(\bm{\mathcal{C}}^{mean}_1,\bm{\mathcal{C}}^{mean}_2,\bm{\mathcal{C}}^{mean}_3,$\\
$\bm{\mathcal{C}}^{mean}_4)$. A class token  ${CLS}\in \mathbb{R}^{1 \times d}$ is concatenated with $\bm{\mathcal{C}}^{mean}\in \mathbb{R}^{4 \times d}$. The concatenated feature is further fed into the next transformer layer to mine the interaction among pathological components. We use a Multi-Layer Perceptron (MLP) head to map the output class token  ${CLS}^{\prime} \in \mathbb{R}^{1 \times d}$ to the final class predictions $P=\mathrm{softmax}\left(\mathrm{MLP}\left({CLS}^{\prime}\right)\right)$. So far, we design a simple yet effective method to achieve the complex biomarker prediction task.
\section{Experiments}
\subsection{Experimental Setup} 
\textbf{Datasets}  We verify the effectiveness of our method on The Caner Genome Atlas (TCGA) CRC dataset and The Clinical Proteomic Tumor Analysis Consortium (CPTAC) CRC dataset, both containing two CRC subtypes, \emph{i.e.}, Microsatellite Stable (MSS) and Microsatellite Instability (MSI) and TCGA also contains another two CRC subtypes, \emph{i.e.}, BRAF mutation and non-BRAF mutation. We use the image preprocessing algorithm in CLAM \cite{lu2021data} to detect the tissue regions in WSIs and crop the tissue regions into non-overlapping 512 × 512 pixel-sized patches at 20× magnification. Due to significant staining variations in the TCGA dataset, we employ the color normalization proposed by Macenko \emph{et al.} \cite{macenko2009method}. Among 482 WSIs in TCGA, 420 of them belong to MSS and 62 are MSI, and 429 of them belong to BRAF mutation and 53 are non-BRAF mutation. We randomly split the TCGA dataset into training and testing sets in a 3:1 ratio, with 10\% of the training set further split into a validation set. Among 105 WSIs in CPTAC, 81 of them belong to MSS and 24 cases are MSI. We randomly split the CPTAC dataset into training, validation, and test sets in a 3:1:1 ratio. \\

\noindent{\textbf{Implementation Details} During training, we evaluate the model on the validation set after every epoch, and save the parameters when it performs the best. Adam is used as our optimizer and the learning rate is $5\times10^{\text{-5}}$ with learning rate annealing. Binary Cross Entropy loss is used as our loss function. The maximum number of training epoch is 32. At last, we measure the performance on the test set. We report the area under the curve (AUC) scores. All experiments are conducted on a single NVIDIA GeForce RTX 3090.}

\subsection{Comparison between PromptBio and Other Methods}
We compare PromptBio with the current state-of-the-art deep MIL models. As shown in Table~\ref{tab1}, our proposed PromptBio significantly outperforms all the compared MIL baselines in AUC. In particular, \emph{e.g.}, over 5\% higher than the second best method TOP* \cite{qu2024rise} and the third best method DSMIL \cite{li2021dual} in TCGA MSI classification. Without pathology text prompts, DSMIL \cite{li2021dual} retains a large number of irrelevant instances, making it challenging to learn genetic biomarker representations. While our PromptBio performs fine-grained pathological component grouping based on pathology text prompts, which effectively extracts instances more related to genetic biomarkers. Though TOP* \cite{qu2024rise} models different pathlogical components guided by pathology language prior knowledge, it neglects the interactions among different components. 
\begin{table}[t]
\centering
\renewcommand\arraystretch{0.8}
\caption{AUC comparison on the TCGA dataset and CPTAC dataset. * denotes that we utilize the framework of the method for our implementation. The best results are marked in bold.  Improvements compared with
the second best results are $\color{right}\textbf{\scriptsize{}highlighted}$.} \label{tab1}
\vspace{-0.5em}
\setlength{\tabcolsep}{1.5mm}
\resizebox{0.85\linewidth}{!}{
\begin{tabular}{lccc}
\toprule[1.5pt]

\textbf{Method} & \textbf{TCGA(MSI)} & \textbf{TCGA(BRAF)} & \textbf{CPTAC(MSI)} \\ \hline
ABMIL \cite{Ilse_Tomczak_Welling_2018}           & 0.8419            & 0.7115             & 0.8000             \\
DSMIL \cite{li2021dual}           & 0.8629            & 0.7310             & 0.8250             \\
CLAM-SB \cite{lu2021data}         & 0.8375            & 0.7522             & 0.8750             \\
CLAM-MB \cite{lu2021data}        & 0.8432            & 0.6373             & 0.8375             \\
TransMIL \cite{shao2021transmil}        & 0.8571            & 0.6708             & 0.8875   
\\
DTFD \cite{zhang2022dtfd}	               &0.8502 	           &0.7467 	           &0.8875 
\\
MHIM-MIL(TransMIL) \cite{tang2023multiple}	&0.7720 	&0.6840 	&0.8750 
\\
TOP* \cite{qu2024rise}	&0.8632  	&0.7176     &0.8875
\\
\rowcolor[HTML]{EFEFEF} 
\textbf{PromptBio}  & \textbf{0.9149}$\color{right}\textbf{\scriptsize{$\uparrow$}5.17}$   & \textbf{0.8025$\color{right}\textbf{\scriptsize{$\uparrow$}5.02}$}    & \textbf{0.9125}$\color{right}\textbf{\scriptsize{$\uparrow$}2.50}$    \\ \bottomrule[1.5pt] 
\end{tabular}}
\vspace{-1em}
\end{table}
\begin{figure}[t]
\centering
\includegraphics[width=\textwidth]{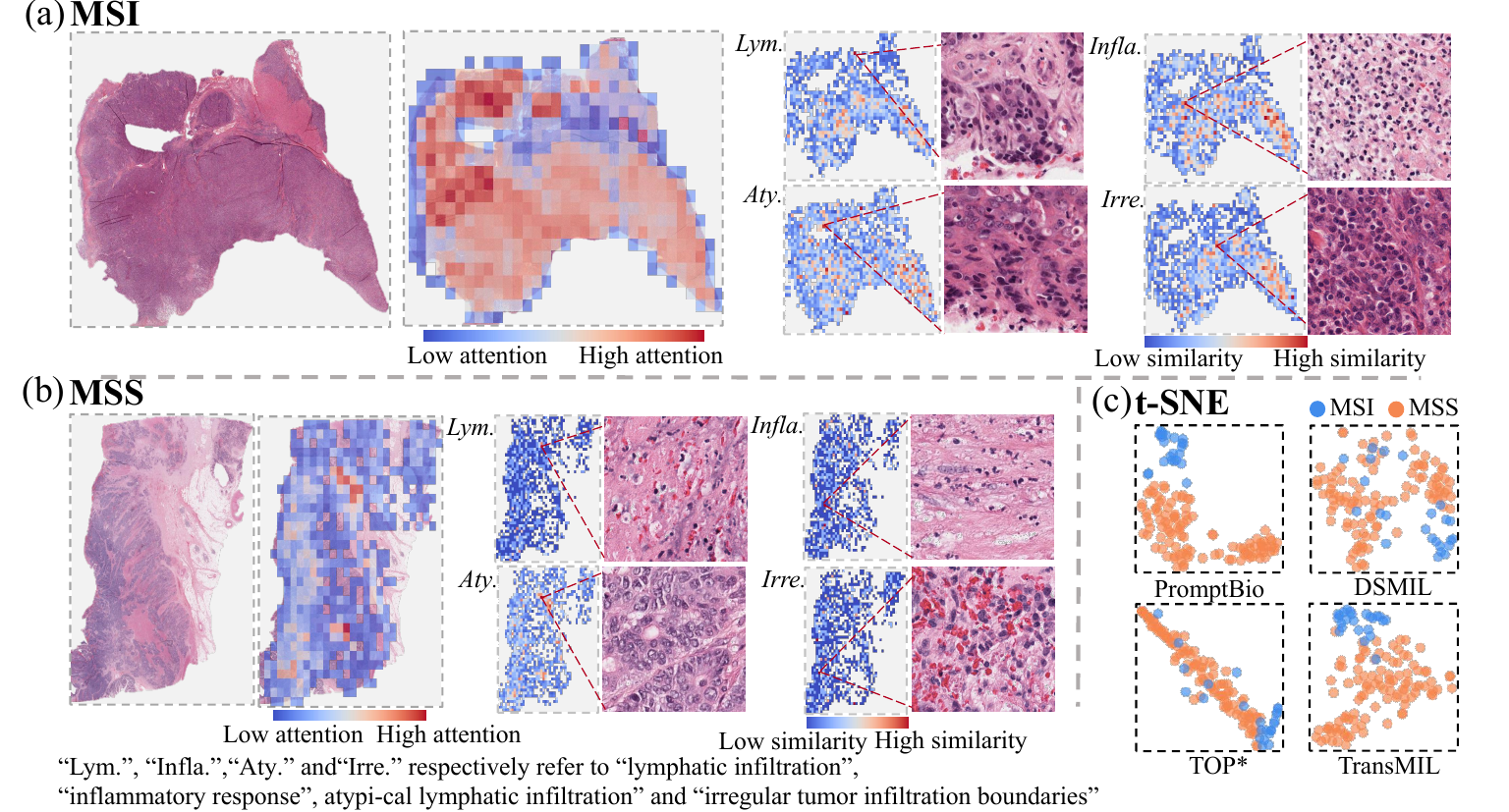}
\vspace{-1em}
\caption{
Attention and t-SNE visualizations. For (a) MSI cancer and (b) MSS cancer, we show attention about contribution of each patch in a pathology image to the class token,  and visualization of similarity between each patch and pathological components. “\emph{Lym.}”, “\emph{Infla.}”,“\emph{Aty.}” and “\emph{Irre.}” respectively refer to “lymphatic infiltration”, “inflammatory response”, atypical lymphatic infiltration” and “irregular tumor infiltration boundaries”. 
(c) t-SNE visualization of bag representations in different methods.
} \label{fig2}
\end{figure}

We further visualize the attention of each patch in pathology images with MSI or MSS cancer and the distribution of pathological components. In Fig.~\ref{fig2} (a), we find that in the pathology image with MSI cancer, large regions receive high attention. The selected patches of inflammatory response and lymphatic infiltration respectively show the accumulation of neutrophils and lymphocytes. While in the pathology image with MSS cancer, regions receiving high attention are much smaller and there are few high-similarity pathology patches of the four conponents. This indicates that prompt-guided fine-grained pathological component grouping can ensure the accuracy of the tumor microenvironment modeling and enhance clinical interpretability. Fig.~\ref{fig2} (c) shows the feature mixtureness of bag representations on testing set. Compared with other methods, our PromptBio yields more discriminative representations.


\subsection{Ablation Study}

Table~\ref{tab2} summarizes the results of ablation study. We evaluate the effectiveness of the proposed modules. Without prompt-guided fine-grained grouping module, the performance drops, \emph{e.g.}, from 91.49\% to 89.65\% in TCGA MSI classification. Without coarse-grained instance selection, the performance also decreases. These indicate that extracted instances based on medical prior knowledge prompting are more related to genetic biomarkers and they can better reflect the tumor microenvironment, while irrelevant pathology instances impeding the learning process of the model. A performance drop without fine-grained pathological component interaction mining module can be observed, \emph{e.g.}, from 91.49\% to 85.97\% in TCGA MSI classification, which shows the importance of exploring the interaction among different pathological components in the tumor microenvironment. Replacing our prompt-guided fine-grained grouping module by K-means clustering algorithm leads to an obvious performance drop. This provides evidence for the effectiveness of the pathology text prompting strategy, enabling a robust modeling of the actual distribution of pathological components. 
Furthermore, to show results are not sensitive \emph{w.r.t.} GPT-4, we change the seed of inference parameters in GPT-4. Though fine-grained text prompt slightly differs, we can still observe similar results (\emph{e.g.}, 91.56\% vs 91.49\% for MSI classification).
\begin{table}[t]
\centering
\renewcommand\arraystretch{0.87}
\caption{Ablation study on TCGA MSI classification, TCGA BRAF classification and CPTAC MSI classification. AUC is reported. Selection, PGG and PCIM denote our coarse-grained instance \textbf{Selection} module, \textbf{P}rompt-\textbf{G}uided fine-grained \textbf{G}rouping module and fine-grained \textbf{P}athological \textbf{C}omponent \textbf{I}nteraction \textbf{M}ining module, respectively. ``K-mean'' means replacing PGG with K-means algorithm. 
``$\triangle$'' means we change the seed of inference parameters in GPT-4 to generate component descriptions.}\label{tab2}
\vspace{-0.5em}
\setlength{\tabcolsep}{1mm}
\resizebox{0.8\linewidth}{!}{
\begin{tabular}{ccc|ccc}
\toprule[1.5pt]
\textbf{Selection} & \textbf{PGG} & \textbf{PCIM} & \textbf{TCGA(MSI)} & \textbf{TCGA(BRAF)} & \textbf{CPTAC(MSI)} \\ \hline
\checkmark                        &                &               & 0.8965             & 0.7310              & 0.8000              \\
                         & \checkmark              & \checkmark             & 0.8902             & 0.7656              & 0.8875              \\
\checkmark                        & \checkmark              & \textbf{}     & 0.8597             & 0.7690              & 0.8625              \\
\checkmark                        & K-means         & \checkmark             & 0.8267             & 0.7266              & 0.7875                   \\

                   \checkmark    &      $\triangle$    & \checkmark             & \textbf{0.9156	}	
		
             & 0.8013              & 0.9125\\
\rowcolor[HTML]{EFEFEF} 
\checkmark                        & \checkmark              & \checkmark             & 0.9149    & \textbf{0.8025}     & \textbf{0.9125}     \\ \bottomrule[1.5pt] 

\end{tabular}
}

\end{table}
\begin{figure}[t]
\centering
\includegraphics[width=0.55\textwidth]{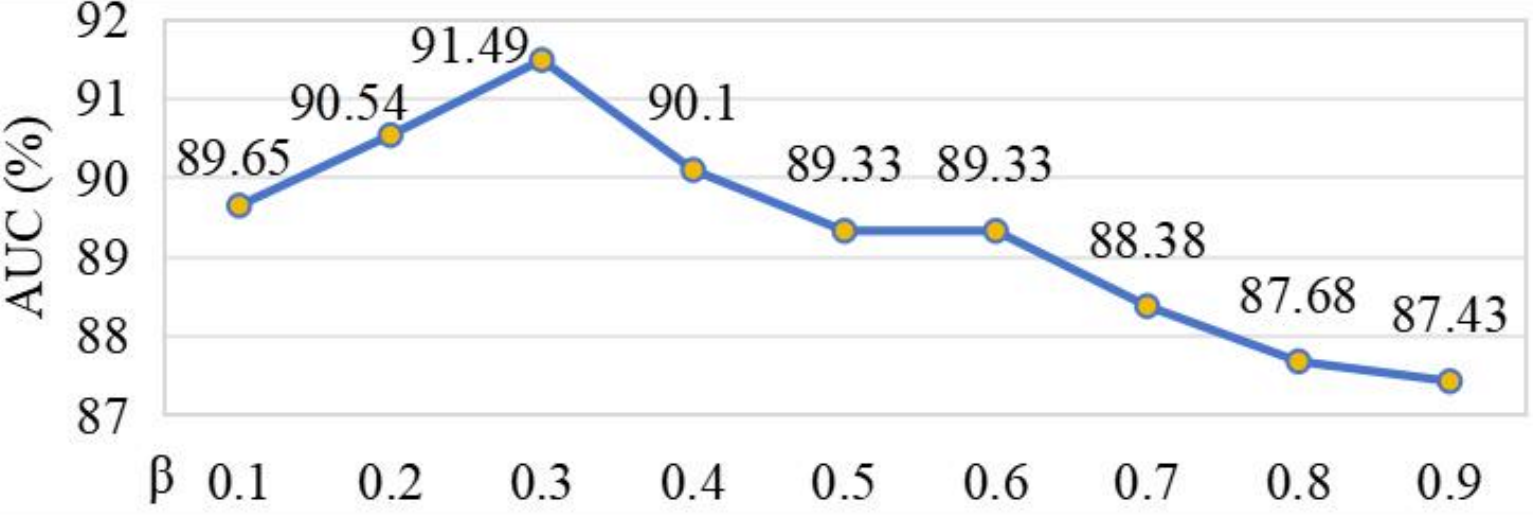}
\vspace{-0.5em}
\caption{Performance changes by varying the selection ratio $\beta$  on TCGA MSI dataset.
} \label{fig3}
\vspace{-1em}
\end{figure}

Fig.~\ref{fig3} shows performance changes by varying the selection ratio $\beta$. It can be seen that the performance is not sensitive within the range of [0.2,0.4]. The model's performance does not always increase as $\beta$ increases, as a larger selection ratio may result in the retention of a large number of irrelevant instances. This observation demonstrates the effectiveness of our prompt-based grouping strategy in extracting instances that are more closely related to genetic biomarkers.

\section{Conclusion}
In this paper, we propose a genetic biomarker prediction method via prompting techniques. We first extract instances related to genetic biomarkers based on medical prior knowledge. Then, guided by pathology text prompts, we group pathology instances from WSIs into fine-grained pathological components and mine their interaction, which enhances clinical interpretability. Experimental results on two CRC cohorts show our model achieves much better performance than other state-of-the-art methods.  

\begin{credits}
\subsubsection{Acknowledgements} This work was supported by the National Natural Science Foundation of China (Grant No. 62101191), Shanghai Natural Science Foundation (Grant No. 21ZR1420800), and the Science and Technology Commission of Shanghai Municipality (Grant No. 22DZ2229004).

\subsubsection{Disclosure of Interests.} The authors have no competing interests to declare that are relevant to the content of this article.
\end{credits}
\bibliographystyle{splncs04}
\bibliography{mybib.bib}





\end{document}


\newcommand{\yan}[1]{ \textcolor{red}{(yan: #1)}  }
\newcommand{\yun}[1]{ \textcolor{blue}{(yun: #1)}  }
\title{Supplementary Material for ``Prompting Whole Slice Image Based Genetic Biomarker Prediction''}
\titlerunning{Factor Space and Spectrum}
%
%
%
%
\author{Anonymous Authors}
\authorrunning{Anonymous Authors}
\institute{Anonymous Institute}

\maketitle

\begin{figure}[h]
\includegraphics[width=\textwidth]{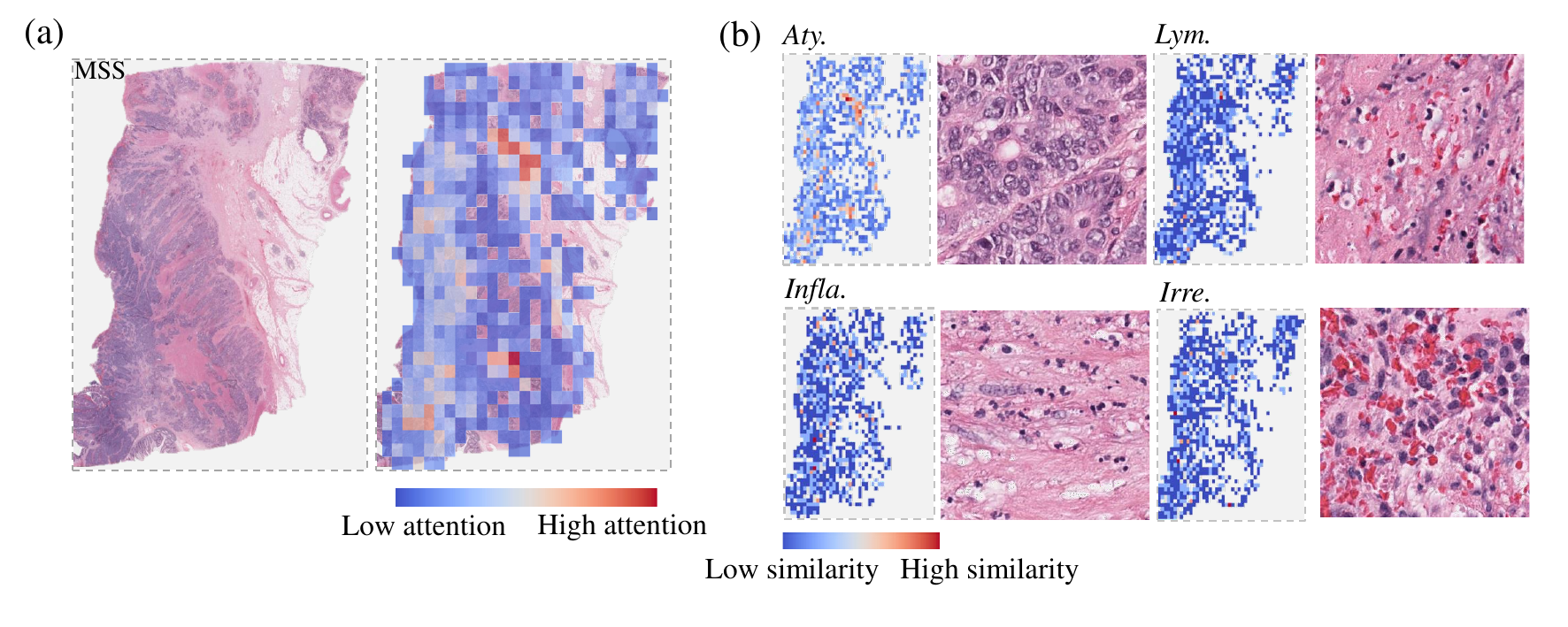}
\caption{PromptBio predicts MSS on TCGA dataset. (a) Attention visualization of a pathology image with MSS cancer. (b) Distribution of each pathological component and the pathology patches that  receive highest similarity with the description. “Lym.”, “Infla.”,“Aty.” and “Irre.” respectively refer to “lymphatic infiltration”, “inflammatory response”, atypical lymphatic infiltration” and “irregular tumor infiltration boundaries”.
} \label{fig2}
\end{figure}
%

